\documentclass[floats, prd, eqnum, showpacs, nofootinbib, 
twocolumn, 
eqsecnum]{revtex4-1}

\usepackage{color,graphicx}
\usepackage{amsfonts}
\usepackage{amssymb}
\usepackage{comment}
\begin{document}

\title{Gravitational lensing by using the $0$th order of affine perturbation series of the deflection angle of a ray near a photon sphere} 
\author{Naoki Tsukamoto${}^{1}$}\email{tsukamoto@rikkyo.ac.jp}

\affiliation{
${}^{1}$Department of General Science and Education, National Institute of Technology, Hachinohe College, Aomori 039-1192, Japan \\
}

\begin{abstract}
The $0$th order of affine perturbation series of the deflection angle of a ray near a photon sphere is more accurate 
than a deflection angle in a strong deflection limit, which is used often, because the later has hidden error terms. 
We investigate gravitational lensing by using $0$th order affine perturbation series of the deflection angle
in a general asymptotically-flat, static, and spherical symmetric spacetime with the photon sphere. 
We apply our formula to Schwarzschild black hole, Reissner-Nordstr\"om black hole, and Ellis-Bronnikov wormhole spacetimes as examples.
By comparing observables by using the deflection angles,
we show that we can ignore the effect of the hidden error terms in the deflection angle in the strong deflection limit on the observables 
in a usual lens configuration with the photon sphere since the hidden error terms are tiny.
On the other hand, in a retro lensing configuration, the deflection angle in the strong-deflection-limit analysis have error of several percent 
and the $0$th order of affine perturbation series of the deflection angle has almost half of the error.
Thus, in the retro lensing configuration, 
we should use the $0$th order of affine perturbation series of the deflection angle rather than the deflection angle in the strong-deflection-limit analysis.
The $0$th order of affine perturbation series of the deflection angle can give a brighter magnification by a dozen percent than 
the one by using the deflection angle in the strong-deflection-limit analysis.
\end{abstract}
\maketitle

\section{Introduction}
Gravitational lensing under a weak-field approximation 
is used to find massive and dark objects~\cite{Schneider_Ehlers_Falco_1992,Schneider_Kochanek_Wambsganss_2006}. 
From the leading term of the deflection angle of a ray reflected by a mass lens in the weak-field approximation, 
we can estimate the mass of the lensing object if a distance to the lensing object is known. 
We would reveal details of the lensing object if we detect the phenomena in a strong gravitational field by the lensing object.~\footnote{  
On this paper, we do not consider rotating lensing objects~\cite{Asada:2000vn,Jusufi:2017mav,Sengo:2022jif}.}

Recently, gravitational waves from black holes have been reported by LIGO Scientific Collaboration and Virgo Collaboration~\cite{Abbott:2016blz}
and the shadows of the candidates of supermassive black holes in the centers of a galaxy M87 and milky way have been reported 
by Event Horizon Telescope Collaboration~\cite{Akiyama:2019cqa,EventHorizonTelescope:2022xnr}.
Investigation on phenomena in strong gravitational fields is important to understand compact objects.

In 1931, Hagihara pointed out that the image of a star at any position can be observed in a Schwarzschild spacetime~\cite{Hagihara_1931} 
because the spacetime has 
a photon sphere~\cite{Ames_1968,Synge:1966okc,Sanchez:1977si,Hasse_Perlick_2002,Yoshino:2019qsh,Claudel:2000yi,Perlick_2004_Living_Rev,Hod:2017xkz,Hod:2017zpi,Koga:2018ybs,Perlick:2021aok}
which is a sphere filled with unstable circular light orbits.
The image due to the rays deflected by the photon sphere around a black hole and other compact objects has been revisited often~\cite{Darwin_1959,Atkinson_1965,Luminet_1979,Ohanian_1987,Nemiroff_1993,Virbhadra:1998dy,Virbhadra_Ellis_2000,Bozza:2001xd,Bozza:2002zj,Virbhadra:2002ju,Perlick:2003vg,Virbhadra:2008ws,Bozza_2010,Tsupko:2017rdo,Bisnovatyi-Kogan:2022ujt,Guerrero:2022qkh,Virbhadra:2022ybp}.
In 1959, Darwin investigated the deflection angle 
of the ray deflected by the photon sphere in the Schwarzschild spacetime~\cite{Darwin_1959}.

Bozza has investigated gravitational lensing in a strong deflection limit $b\rightarrow b_\mathrm{m}+0$, where $b$ is the impact parameter  of the ray and 
$b_\mathrm{m}$ is a critical impact parameter, in a general asymptotically-flat, spherical symmetric spacetime with the photon sphere~\cite{Bozza:2002zj}.  
Bozza has expressed the deflection angle $\alpha$ of a ray reflected by the photon sphere as
\begin{eqnarray}\label{eq:alp11}
\alpha=-\bar{a}\log\left( \frac{b}{b_\mathrm{m}}-1 \right) +\bar{b}, 
\end{eqnarray}
where $\bar{a}$ and $\bar{b}$ can be calculated by using the metric of the spacetime.~\footnote{
In Ref.~\cite{Bozza:2002zj}, the order of the error of Eq.~(\ref{eq:alp11}) is estimated as  
$O\left( \frac{b}{b_\mathrm{m}}-1 \right)$.
In Refs.~\cite{Tsukamoto:2016jzh,Tsukamoto:2016qro},
Tsukamoto claims that the order of the error should read as $O\left(\left( \frac{b}{b_\mathrm{m}}-1 \right) \log \left( \frac{b}{b_\mathrm{m}}-1 \right) \right)$.  
Iyer and Petters~\cite{Iyer:2006cn} and Tsukamoto~\cite{Tsukamoto:2022tmm}
discuss hidden error terms in the deflection angle~(\ref{eq:alp11}).}
In many spacetimes, $\bar{a}$ is obtained as analytical forms while
$\bar{b}$ usually is calculated numerically. 
Analytic forms of $\bar{a}$ and $\bar{b}$ have been obtained 
only in simple spacetimes such as the Schwarzschild spacetime~\cite{Bozza:2001xd,Bozza:2002zj},
higher dimensional black hole spacetimes~\cite{Eiroa:2004gh,Tsukamoto:2014dta},
charged black hole spacetimes~\cite{Tsukamoto:2016oca,Tsukamoto:2016jzh,Badia:2017art},
rotating black hole spacetimes~\cite{Hsieh:2021scb},
and wormhole spacetimes~\cite{Tsukamoto:2016qro,Tsukamoto:2016zdu}.
The analysis in the strong deflection limit has been extended 
and applied to various astrophysical situations~\cite{Tsukamoto:2016zdu,Shaikh:2019jfr,Shaikh:2019itn,Tsukamoto:2020uay,Tsukamoto:2020iez,Paul:2020ufc,Bozza:2002af,Eiroa:2002mk,Petters:2002fa,Eiroa:2003jf,Bozza:2004kq,Bozza:2005tg,Bozza:2006sn,Bozza:2006nm,Iyer:2006cn,Bozza:2007gt,Tsukamoto:2016qro,Ishihara:2016sfv,Tsukamoto:2016oca,Tsukamoto:2016jzh,Tsukamoto:2017edq,Hsieh:2021scb,Aldi:2016ntn,Tsukamoto:2020bjm,Takizawa:2021gdp,Tsukamoto:2021caq,Aratore:2021usi,Bisnovatyi-Kogan:2022ujt,Tsupko:2022kwi,Ghosh:2022mka}.

Iyer and Petters have investigated affine perturbation series of the deflection angle near the photon sphere in the Schwarzschild spacetime in the following form:  
\begin{eqnarray}\label{eq:affine}
\alpha
&=& ( \sigma_0 +\sigma_1 b_\mathrm{p}+\sigma_2 b_\mathrm{p}^2 +\sigma_3 b_\mathrm{p}^3 +\cdots  ) 
\log \left( \frac{\lambda_0}{b_\mathrm{p}} \right) \nonumber\\
&&+\rho_0+\rho_1 b_\mathrm{p}+\rho_2 b_\mathrm{p}^2 +\rho_3 b_\mathrm{p}^3 +\cdots, 
\end{eqnarray}
where $b_\mathrm{p}$ is defined by
\begin{equation}
b_\mathrm{p} \equiv 1-\frac{b_\mathrm{m}}{b},
\end{equation}
and $\lambda_0$, $\sigma_0$, $\sigma_1$, $\sigma_2$, $\sigma_3$, $\rho_0$, $\rho_1$, $\rho_2$, and $\rho_3$ are constant,
and they have found the $0$th order of the affine perturbation series 
\begin{eqnarray}\label{eq:affine0}
\alpha
=\sigma_0 \log \left( \frac{\lambda_0}{b_\mathrm{p}} \right) +\rho_0
\end{eqnarray}
is more accurate than the deflection angle by Darwin~\cite{Iyer:2006cn}.
Tsukamoto has investigated the affine perturbation series of the deflection angle in the Reissner-Nordstr\"om black hole spacetime
and has confirmed the $0$th order of affine perturbation series (\ref{eq:affine0}) is more accurate than the form of Eq.~(\ref{eq:alp11}).

How much does the difference of the deflection angles (\ref{eq:alp11}) and (\ref{eq:affine0}) affect observables in gravitational lensing?
To answer this question, we investigate gravitational lensing in a general asymptotically-flat, static, and spherical symmetric spacetime with the photon sphere 
by using deflection angle in a form 
\begin{eqnarray}\label{eq:alp12}
\alpha=-\bar{a}\log\left( 1-\frac{b_\mathrm{m}}{b} \right) +\bar{b},
\end{eqnarray}
which is the same as the $0$th order of affine perturbation series~(\ref{eq:affine0}) 
with the relations
\begin{eqnarray}
\bar{a}= \sigma_0  
\end{eqnarray}
and 
\begin{eqnarray}
\bar{b}= \sigma_0 \log \lambda_0 + \rho_0
\end{eqnarray}
in a usual lens configuration and a retro lensing configuration. 

This paper is organized as follows. 
We investigate the $0$th order of affine perturbation series of the deflection angle~(\ref{eq:alp12}) in Sec. II 
and we consider the Schwarzschild black hole, Reissner-Nordstr\"om black hole, and the Ellis-Bronnikov wormhole spacetimes in Sec.~III.
We investigate gravitational lensing by the photon sphere in a usual lens configuration in Sec.~IV and in a retro lens configuration in Sec.~V.  
We conclude and discuss in Sec.~VI.
We review gravitational lensing under weak-field approximations in the usual lens configuration in appendix~A.
We use the units in which the light speed and Newton's constant are unity.

\section{$0$th order of affine perturbation series of the deflection angle~(\ref{eq:alp12})}
In this section, we investigate the $0$th order of affine perturbation series of the deflection angle~(\ref{eq:alp12})
in a general, asymptotically flat, static, and spherically symmetric spacetime
with a metric 
\begin{eqnarray}
ds^2=-A(r)dt^2+B(r)dr^2+C(r)(d\vartheta^2+\sin^2 \vartheta d\varphi^2) \nonumber\\
\end{eqnarray}
and with time translational and axial Killing vectors $t^\mu\partial_\mu=\partial_t$ and $\varphi^\mu\partial_\mu=\partial_\varphi$, respectively.

We assume a photon sphere at $r=r_\mathrm{m}$ which is the largest positive solution of $D(r)=0$, where $D(r)$ is defined by
\begin{eqnarray}
D(r)\equiv \frac{C^\prime (r)}{C(r)}-\frac{A^\prime (r)}{A(r)},
\end{eqnarray}
where the prime denotes a differentiation with respect to $r$.
We also assume that 
$A(r)$, $B(r)$, and $C(r)$ satisfy an asymptotically-flat condition
\begin{eqnarray}
\lim_{r\rightarrow \infty} A(r)=\lim_{r\rightarrow \infty} B(r)= \lim_{r\rightarrow \infty} \frac{C(r)}{r^2}=1,
\end{eqnarray}
and that $A(r)$, $B(r)$, and $C(r)$ are positive and finite for $r>r_\mathrm{m}$.
We assume $\vartheta=\pi/2$ without loss of generality because of spherical symmetry.

The trajectory of the ray is expressed by
\begin{eqnarray}\label{eq:tra}
-A(r)\dot{t}^2+B(r)\dot{r}^2+C(r)\dot{\varphi}^2=0,
\end{eqnarray}
where the dot denotes a differentiation with respect to an affine parameter along the trajectory.
Conserved energy $E\equiv -g_{\mu\nu}t^\mu\dot{x}^\nu=A(r)\dot{t}$ and angular momentum $L\equiv g_{\mu\nu}\varphi^\mu\dot{x}^\nu=C(r)\dot{\varphi}$ of the ray 
are constant along the trajectory
and the impact parameter of the ray is defined by $b\equiv L/E$.
For simplicity, we assume that the impact parameter is positive in this section.
The trajectory can be rewritten as 
\begin{eqnarray}
\dot{r}^2+V(r)=0,
\end{eqnarray}
where $V(r)$ is an effective potential defined by
\begin{eqnarray}
V(r)\equiv \frac{L^2 R(r)}{B(r)C(r)},
\end{eqnarray}
where $R(r)$ is defined by
\begin{eqnarray}
R(r)\equiv \frac{C(r)}{A(r)b^2}-1.
\end{eqnarray}
We assume that the effective potential is negative $V(r)<0$ for $r_\mathrm{m}<r<\infty$ so that the ray reaches to the photon sphere from spatial infinity. 

We concentrate on a scatter case since we are interested in gravitational lensing. 
In this case, the ray is scattered at a closest distance $r=r_0>r_\mathrm{m}$.
Equation~(\ref{eq:tra}) gives
\begin{eqnarray}\label{eq:tra2}
A_0\dot{t}_0^2=C_0\dot{\varphi}_0^2
\end{eqnarray}
at the closest distance $r=r_0$.
Here and hereafter, quantities with the subscript $0$ denotes the quantities at $r=r_0$.
From Eq.~(\ref{eq:tra2}), the positive impact parameter is expressed by
\begin{eqnarray}
b=b(r_0)=\frac{L}{E}=\frac{C_0\dot{\varphi}_0}{A_0\dot{t}_0}=\sqrt{\frac{C_0}{A_0}}
\end{eqnarray}
and $R$ can be rewritten as 
\begin{eqnarray}
R=R(r,r_0)=\frac{A_0C(r)}{A(r)C_0}-1.
\end{eqnarray}
At the closest distance, we obtain
\begin{eqnarray}
R_0=V_0=0
\end{eqnarray}
and 
\begin{eqnarray}
R^\prime_0&=&\frac{D_0}{C_0^2}, \\
V^\prime_0&=&\frac{L^2}{B_0 C_0}  R^\prime_0, \\
V^{\prime \prime}_0&=&\left(2\frac{L^2}{B_0 C_0} \right)^\prime R^\prime_0+\frac{L^2 }{B_0 C_0} R^{\prime \prime}_0.
\end{eqnarray}
In a strong deflection limit $r_0 \rightarrow r_\mathrm{m}+0$ or $b \rightarrow b_\mathrm{m}+0$, where the critical impact parameter $b_\mathrm{m}$ is defined by
\begin{eqnarray}
b_\mathrm{m}\equiv \lim_{r_0 \rightarrow r_\mathrm{m}+0} \sqrt{\frac{C_0}{A_0}},
\end{eqnarray}
we obtain 
\begin{eqnarray}
D_\mathrm{m}&\equiv&\lim_{r_0 \rightarrow r_\mathrm{m}+0} D_0=\lim_{r \rightarrow r_\mathrm{m}+0} D(r)=0, \\
R^\prime_\mathrm{m}&\equiv&\lim_{r_0 \rightarrow r_\mathrm{m}+0} R^\prime_0=0, \\
V^\prime_\mathrm{m}&\equiv&\lim_{r_0 \rightarrow r_\mathrm{m}+0} V^\prime_0=0. 
\end{eqnarray}

We can rewrite Eq.~(\ref{eq:tra}) as  
\begin{eqnarray}
\left( \frac{dr}{d\varphi} \right)^2 =\frac{R(r,r_0)C(r)}{B(r)}
\end{eqnarray}
and we obtain the deflection angle $\alpha(r_0)$ of the ray as 
\begin{eqnarray}\label{eq:alpha}
\alpha(r_0)=I(r_0)-\pi,
\end{eqnarray}
where $I(r_0)$ is defined by 
\begin{eqnarray}
I(r_0)\equiv 2\int^\infty_{r_0} \frac{dr}{\sqrt{\frac{R(r,r_0)C(r)}{B(r)}}}.
\end{eqnarray}
We change the radial coordinate $r$ to a variable $z$ defined by
\begin{eqnarray}
z\equiv 1-\frac{r_0}{r}
\end{eqnarray}
and we obtain $I(r_0)$ as 
\begin{eqnarray}
I(r_0)=\int^1_0 f(z,r_0) dz,
\end{eqnarray}
where $f(z,r_0)$ is defined by
\begin{eqnarray}
f(z,r_0) \equiv \frac{2r_0}{\sqrt{G(z,r_0)}},
\end{eqnarray}
where $G(z,r_0)$ is defined as 
\begin{eqnarray}\label{eq:G}
G(z,r_0) \equiv R(r(z),r_0)\frac{C(r(z))}{B(r(z))}(1-z)^4.
\end{eqnarray}
By using the expansions of a function $F(r(z))$ and its inverse $1/F(r(z))$ in the power of $z$, which are expressed by  
\begin{equation}
F=F_0+F^\prime_0 r_0 z + \left(\frac{1}{2}F^{\prime \prime}_0 r^2_0+F^\prime_0 r_0 \right)z^2+O\left( z^3 \right)
\end{equation}
and 
\begin{eqnarray}
\frac{1}{F}
&=&\frac{1}{F_0}-\frac{F_0^\prime r_0}{F_0^2} z \nonumber\\
&&+\left( \frac{r_0^2 F_0^{\prime 2}}{F_0^3}-\frac{r_0 F_0^\prime}{F_0^2}-\frac{r_0^2 F_0^{\prime \prime}}{2F_0^2} \right)z^2  +O\left( z^3 \right), \qquad 
\end{eqnarray}
respectively, we obtain the expansion of $R(r(z),r_0)$ in the power of $z$ as 
\begin{eqnarray}\label{eq:Rexpa}
R
&=&D_0 r_0 z+ \left[ \frac{r_0}{2} \left( \frac{C_0^{\prime \prime}}{C_0}-\frac{A_0^{\prime \prime}}{A_0} \right) \right. \nonumber\\
&&+ \left. \left( 1-\frac{A_0^\prime r_0}{A_0}\right) D_0 \right] r_0 z^2+ O\left( z^3 \right).
\end{eqnarray}
From Eqs.~(\ref{eq:G})-(\ref{eq:Rexpa}), $G(z,r_0)$ can be expanded in the power of $z$ as 
\begin{eqnarray}
G(z,r_0) =c_1(r_0) z+ c_2(r_0) z^2 +O\left( z^3 \right),
\end{eqnarray}
where $c_1(r_0)$ and $c_2(r_0)$ are obtained as 
\begin{eqnarray}
c_1(r_0)= \frac{C_0 D_0 r_0}{B_0}
\end{eqnarray}
and 
\begin{eqnarray}
c_2(r_0)
&=& \frac{C_0 r_0}{B_0} \left\{ D_0 \left[ \left( D_0-\frac{B_0^\prime}{B_0} \right)r_0-3 \right] \right. \nonumber\\
&&\left. +\frac{r_0}{2} \left(\frac{C_0^{\prime \prime}}{C_0}-\frac{A_0^{\prime \prime}}{A_0} \right)  \right\}, 
\end{eqnarray}
respectively.
In the strong deflection limit $r_0 \rightarrow r_\mathrm{m}+0$, we obtain
\begin{eqnarray}
c_1(r_\mathrm{m})=0 
\end{eqnarray}
and 
\begin{eqnarray}
c_2(r_\mathrm{m})=\frac{C_\mathrm{m} r_\mathrm{m}^2}{2 B_\mathrm{m}} D_\mathrm{m}^\prime, 
\end{eqnarray}
where we define
\begin{eqnarray}
A_\mathrm{m}&\equiv& \lim_{r_0 \rightarrow r_\mathrm{m}+0} A_0=\lim_{r \rightarrow r_\mathrm{m}+0} A, \\
B_\mathrm{m}&\equiv& \lim_{r_0 \rightarrow r_\mathrm{m}+0} B_0=\lim_{r \rightarrow r_\mathrm{m}+0} B, \\
C_\mathrm{m}&\equiv& \lim_{r_0 \rightarrow r_\mathrm{m}+0} C_0=\lim_{r \rightarrow r_\mathrm{m}+0} C, \\
D_\mathrm{m}^\prime&\equiv& \lim_{r_0 \rightarrow r_\mathrm{m}+0} D_0^\prime=\lim_{r \rightarrow r_\mathrm{m}+0} D^\prime=\frac{C_\mathrm{m}^{\prime \prime}}{C_\mathrm{m}}-\frac{A_\mathrm{m}^{\prime \prime}}{A_\mathrm{m}} \nonumber\\
\end{eqnarray}
and then we get
\begin{eqnarray}
G(z,r_\mathrm{m})=c_2(r_\mathrm{m})z^2+O(z^3).
\end{eqnarray}
We assume that $D_\mathrm{m}^\prime$ does not vanish. 
Under the assumption, the term $I(r_0)$ diverges logarithmically 
in the strong deflection limit $r_0 \rightarrow r_\mathrm{m}+0$.~\footnote{The cases of $D_\mathrm{m}^\prime=0$ are discussed in Refs.~\cite{Tsukamoto:2020uay,Tsukamoto:2020iez,Chiba:2017nml}.}
We define the divergent part $I_\mathrm{D}(r_0)$ of the term $I(r_0)$ as 
\begin{eqnarray}
I_\mathrm{D}(r_0)
&\equiv& \int^1_0 f_\mathrm{D} (z,r_0) dz \nonumber \\
&=&\frac{4r_0}{\sqrt{c_2(r_0)}} \log \frac{\sqrt{c_2(r_0)}+\sqrt{c_1(r_0)+c_2(r_0)}}{\sqrt{c_1(r_0)}}, \nonumber\\
\end{eqnarray}
where $f_\mathrm{D}(z,r_0)$ is defined as
\begin{eqnarray}
f_\mathrm{D}(z,r_0)\equiv \frac{2r_0}{\sqrt{c_1(r_0)z+c_2(r_0)z^2}}.
\end{eqnarray}
We expand $c_1(r_0)$ and $b(r_0)$ in powers of $r_0-r_\mathrm{m}$ as 
\begin{eqnarray}
c_1(r_0)=\frac{C_\mathrm{m} r_\mathrm{m} D^\prime_\mathrm{m}}{B_\mathrm{m}} (r_0-r_\mathrm{m})+O\left((r_0-r_\mathrm{m})^2\right)
\end{eqnarray}
and 
\begin{eqnarray}
b(r_0)=b_\mathrm{m} +\frac{1}{4}\sqrt{\frac{C_\mathrm{m}}{A_\mathrm{m}}}D_\mathrm{m}^\prime (r_0-r_\mathrm{m})^2 +O\left((r_0-r_\mathrm{m})^3\right), \nonumber\\
\end{eqnarray}
respectively.
Therefore, we obtain the relation, in the strong deflection limit $r_0 \rightarrow r_\mathrm{m}+0$ or $b \rightarrow b_\mathrm{m}+0$,
\begin{eqnarray}
&&\lim_{r_0 \rightarrow r_\mathrm{m} +0} c_1(r_0) \nonumber\\
&&=\lim_{b \rightarrow b_\mathrm{m} +0} \frac{2C_\mathrm{m}r_\mathrm{m}\sqrt{D_\mathrm{m}^\prime}}{B_\mathrm{m}} \left( \frac{b}{b_\mathrm{m}}-1 \right)^\frac{1}{2} \nonumber\\
&&=\lim_{b \rightarrow b_\mathrm{m} +0} \frac{2C_\mathrm{m}r_\mathrm{m}\sqrt{D_\mathrm{m}^\prime}}{B_\mathrm{m}} \left( 1-\frac{b_\mathrm{m}}{b} \right)^\frac{1}{2}.
\end{eqnarray}
By using the relation, we express the divergent part $I_\mathrm{D}(b)$ in the strong deflection limit $b \rightarrow b_\mathrm{m}+0$ as 
\begin{eqnarray}
I_\mathrm{D}(b)
&=&-\frac{r_\mathrm{m}}{\sqrt{c_2(r_\mathrm{m})}} \log \left( 1-\frac{b_\mathrm{m}}{b} \right) +\frac{r_\mathrm{m}}{\sqrt{c_2(r_\mathrm{m})}} \log r_\mathrm{m}^2 D_\mathrm{m}^\prime \nonumber\\
&&+O\left( \left( 1-\frac{b_\mathrm{m}}{b} \right) \log \left( 1-\frac{b_\mathrm{m}}{b} \right) \right).
\end{eqnarray}

We define the regular part $I_\mathrm{R}(r_0)$ of the term $I(r_0)$ as 
\begin{eqnarray}
I_\mathrm{R}(r_0) \equiv \int^1_0 f_\mathrm{R} (z,r_0) dz, 
\end{eqnarray}
where $f_\mathrm{R} (z,r_0)$ is defined as
\begin{eqnarray}
f_\mathrm{R} (z,r_0) \equiv f(z,r_0)-f_\mathrm{D}(z,r_0),
\end{eqnarray}
and expand it in power of $r_0 -r_\mathrm{m}$ 
and we only consider the first term in which we are interested.
Then, we get  
\begin{eqnarray}
I_\mathrm{R}(r_0)
&=& \int^1_0 f_\mathrm{R} (z,r_\mathrm{m}) dz,  \nonumber\\
&&+O\left( \left( 1-\frac{r_\mathrm{m}}{r_0} \right) \log \left( 1-\frac{r_\mathrm{m}}{r_0} \right) \right)
\end{eqnarray}
or
\begin{eqnarray}
I_\mathrm{R}(b)
&=& \int^1_0 f_\mathrm{R} (z,b_\mathrm{m}) dz,  \nonumber\\
&&+O\left( \left( 1-\frac{b_\mathrm{m}}{b} \right) \log \left( 1-\frac{b_\mathrm{m}}{b} \right) \right).
\end{eqnarray}
From $I=I_\mathrm{D}+I_\mathrm{R}$, we obtain the deflection angle in the strong deflection limit $b\rightarrow b_\mathrm{m}+0$ as 
\begin{eqnarray}
\alpha(b)
&=&-\bar{a} \log \left( 1-\frac{b_\mathrm{m}}{b} \right) +\bar{b} \nonumber\\
&&+O\left( \left( 1-\frac{b_\mathrm{m}}{b} \right) \log \left( 1-\frac{b_\mathrm{m}}{b} \right) \right),
\end{eqnarray}
where $\bar{a}$ and $\bar{b}$ are given by
\begin{eqnarray}
\bar{a}=\sqrt{\frac{2B_\mathrm{m} A_\mathrm{m}}{C^{\prime \prime}_\mathrm{m} A_\mathrm{m}- C_\mathrm{m} A^{\prime \prime}_\mathrm{m}}}
\end{eqnarray}
and 
\begin{eqnarray}
\bar{b}=\bar{a} \log \left[ r^2_\mathrm{m} \left( \frac{C^{\prime \prime}_\mathrm{m}}{C_\mathrm{m}}- \frac{A^{\prime \prime}_\mathrm{m}}{A_\mathrm{m}} \right) \right] +I_\mathrm{R} (b_\mathrm{m}) -\pi, 
\end{eqnarray}
respectively.

\section{Examples of deflection angles}
In this section, we apply the formulas in the previous section to the Schwarzschild black hole, 
Reissner-Nordstr\"om black hole and Ellis-Bronnikov wormhole spacetimes.
We obtain $\bar{a}$ and $\bar{b}$ in the deflection angles~(\ref{eq:alp11}) and~(\ref{eq:alp12}) and 
we show the percent errors of the deflection angles~(\ref{eq:alp11}) and~(\ref{eq:alp12}) defined by
\begin{equation}
\frac{\alpha \: \mathrm{of}\: \mathrm{Eq}.(2.20)-\alpha \: \mathrm{of}\: \mathrm{Eq}.(1.1)}{\alpha \: \mathrm{of} \: \mathrm{Eq}.(2.20)} \times 100
\end{equation}
and 
\begin{equation}
\frac{\alpha \: \mathrm{of}\: \mathrm{Eq}.(2.20)-\alpha \: \mathrm{of}\: \mathrm{Eq}.(1.5)}{\alpha \: \mathrm{of} \: \mathrm{Eq}.(2.20)} \times 100,
\end{equation}
respectively.

\subsection{Schwarzschild black hole}
In the Schwarzschild spacetime with a mass $M$, the functions $A(r)$, $B(r)$, and $C(r)$ are given by 
\begin{equation}
A(r)=1-\frac{2M}{r},
\end{equation}
\begin{equation}
B(r)=\frac{1}{1-\frac{2M}{r}},
\end{equation}
and 
\begin{equation}
C(r)=r^2,
\end{equation}
respectively.

\begin{figure}[htbp]
\begin{center}
\includegraphics[width=75mm]{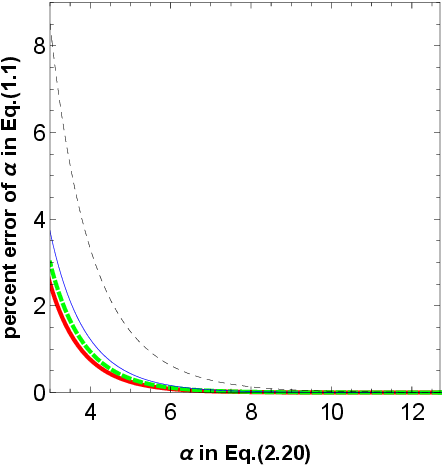}\\
\includegraphics[width=75mm]{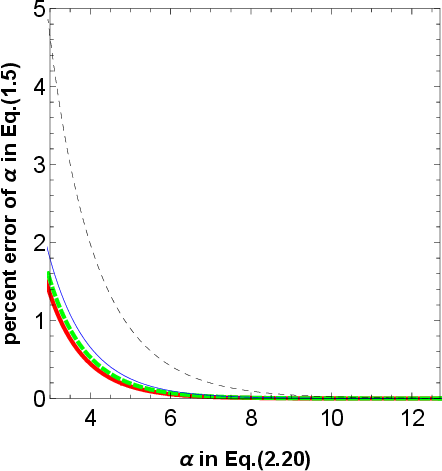}
\end{center}
\caption{The percent errors of deflection angles in the Schwarzschild and Reissner-Nordstr\"om black hole spacetimes. 
The percent errors of the deflection angles in Eqs.~(\ref{eq:alp11}) and (\ref{eq:alp12}) against the deflection angle in Eq.~(\ref{eq:alpha})
are shown in the upper and lower panels, respectively.
Wide solid (red), wide dashed (green), narrow solid (blue), and narrow dashed (black) curves show the percent errors 
in the cases of $Q/M=0$, $0.6$, $0.8$, and $1$, respectively.}
\label{fig:error}
\end{figure}

The critical impact parameter is given by $b_\mathrm{m}= 3\sqrt{3}M$ and 
the photon sphere is at $r=r_\mathrm{m}=3M$.
As obtained in Refs.~\cite{Bozza:2001xd,Bozza:2002zj},
coefficients 
$\bar{a}$ and $\bar{b}$ of the deflection angles in the strong deflection limit 
are obtained as
\begin{equation}
\bar{a}=1
\end{equation}
and 
\begin{equation}
\bar{b}=\log 216(7-4\sqrt{3}) -\pi,
\end{equation}
respectively.
The percent errors of the deflection angles~(\ref{eq:alp11}) and~(\ref{eq:alp12}) are shown in Fig.~\ref{fig:error}.

\subsection{Reissner-Nordstr\"om black hole}
A Reissner-Nordstr\"om black hole is often considered as the simplest extension of the Schwarzschild black hole.
Gravitational lensing~\cite{Eiroa:2002mk,Bozza:2002zj,Eiroa:2003jf,Bin-Nun:2010exl,Bin-Nun:2010lws,Tsukamoto:2016oca,Tsukamoto:2016jzh}, 
shadow~\cite{deVries:2000,Takahashi:2005hy,Zakharov:2014lqa,Akiyama:2019cqa,Akiyama:2019eap,Kocherlakota:2021dcv},
and time delay~\cite{Sereno:2003nd} by the Reissner-Nordstr\"om black hole have been investigated.
 
Eiroa has considered gravitational lensing by the Reissner-Nordstr\"om black hole 
in the strong deflection limit $b \rightarrow b_\mathrm{m}+0$ in numerical~\cite{Eiroa:2002mk}.
Coefficients $\bar{a}$ in an analytical form and $\bar{b}$ in numerical have been obtained by Bozza~\cite{Bozza:2002zj}.
The analytical forms of $\bar{a}$ and $\bar{b}$ have been obtained in Refs.~\cite{Tsukamoto:2016oca,Tsukamoto:2016jzh}.

In the Reissner-Nordstr\"om black hole spacetime for $0\leq Q^2/M^2\leq 1$, where $Q$ is an electrical charge,
\footnote{Gravitational lensing and shadow in overcharged cases for $1<Q^2/M^2$ have been investigated 
in Refs.~\cite{Chiba:2017nml,Shaikh:2019itn,Tsukamoto:2021fsz,Tsukamoto:2021lpm,Tsukamoto:2020iez}. 
} 
the functions $A(r)$, $B(r)$, and $C(r)$ are given by 
\begin{equation}
A(r)=1-\frac{2M}{r}+\frac{Q^2}{r^2},
\end{equation}
\begin{equation}
B(r)=\frac{1}{1-\frac{2M}{r}+\frac{Q^2}{r^2}},
\end{equation}
and 
\begin{equation}
C(r)=r^2,
\end{equation}
respectively.

We obtain $r_\mathrm{m}$ as  
\begin{equation}
r_\mathrm{m}=\frac{3M+\sqrt{9M^2-8Q^2}}{2}
\end{equation}
and $b_\mathrm{m}$ as
\begin{equation}
b_\mathrm{m}=\frac{r_\mathrm{m}^2}{\sqrt{Mr_\mathrm{m}-Q^2}}.
\end{equation}
The coefficients $\bar{a}$ and $\bar{b}$ of the deflection angles in the strong deflection limit are given by
\begin{equation}
\bar{a}=\frac{r_\mathrm{m}}{\sqrt{3Mr_\mathrm{m}-4Q^2}}
\end{equation}
and 
\begin{eqnarray}
\bar{b}
&=&\bar{a} \log \left[ \frac{8(3Mr_\mathrm{m}-4Q^2)^3}{M^2 r_\mathrm{m}^2 (Mr_\mathrm{m}-Q^2)^2} \right. \nonumber\\ 
&&\left. \times \left( 2\sqrt{Mr_\mathrm{m}-Q^2}-\sqrt{3Mr_\mathrm{m}-4Q^2} \right)^2 \right] -\pi, \nonumber\\ 
\end{eqnarray}
respectively.
We show the percent errors of the deflection angles~(\ref{eq:alp11}) and~(\ref{eq:alp12}) in Fig.~\ref{fig:error}.

\subsection{An Ellis-Bronnikov wormhole}
An Ellis-Bronnikov wormhole is the solution of Einstein equations with a phantom scalar field~\cite{Ellis:1973yv,Bronnikov:1973fh}.
The deflection angle in the Ellis-Bronnikov wormhole spacetime has investigated by~Chetouani and Cl\'{e}ment~\cite{Chetouani_Clement_1984}
and it has been revisited by several authors~\cite{Nandi:2006ds,Muller:2008zza,Bhattacharya:2010zzb,Gibbons:2011rh,Nakajima:2012pu,Tsukamoto:2012xs,Tsukamoto:2017edq,Jusufi:2017gyu}.
The visual appearance of the wormhole~\cite{Muller:2004dq}
and images due to the 
photon sphere~\cite{Perlick:2003vg,Nandi:2006ds,Tsukamoto:2012xs,Perlick:2015vta,Ohgami:2015nra,Ohgami:2016iqm,Nandi:2016ccg,Nandi:2016uzg,Tsukamoto:2016zdu,Tsukamoto:2016jzh,Tsukamoto:2017edq,Shaikh:2019jfr}
have been investigated.

We cannot apply directly Bozza's method~\cite{Bozza:2002zj} to an ultrastatic spacetime with a time translational Killing vector with a constant norm 
such as the Ellis-Bronnikov wormhole spacetime~\footnote{We can apply indirectly Bozza's method to the ultrastatic Ellis-Bronnikov wormhole~\cite{Bhattacharya:2019kkb}.}. 
An extended method for the ultrastatic spacetime has been investigated 
and the deflection angle in the strong deflection limit in the Ellis-Bronnikov wormhole spacetime has been calculated in Refs.~\cite{Tsukamoto:2016zdu,Tsukamoto:2016qro,Tsukamoto:2016jzh}. 

A line element in the Ellis-Bronnikov wormhole spacetime is given by  
\begin{equation}
ds^2=-dt^2+dl^2+(l^2+a^2)(d\vartheta^2+\sin^2 \vartheta d\varphi^2),
\end{equation}
where $a$ is a positive constant. 
We cannot apply formulas in Sec.~II in a radial coordinate $l$ since the photon sphere is at $l=0$.
We use a radial coordinate $r$ defined by $r\equiv l+p$, where $p$ is a positive constant, 
so that the photon sphere is at $r=r_\mathrm{m}=p>0$. 
Under the radial coordinate $r$, we get 
\begin{equation}
A(r)=1,
\end{equation}
\begin{equation}
B(r)=1,
\end{equation}
and 
\begin{equation}
C(r)=(r-p)^2+a^2.
\end{equation}

The critical impact parameter is given by $b_\mathrm{m}=a$
and coefficients of the deflection angles in the strong deflection limit are obtained as~\cite{Tsukamoto:2016qro} 
\begin{equation}
\bar{a}=1
\end{equation}
and 
\begin{equation}
\bar{b}=3 \log 2 -\pi.
\end{equation}
\begin{figure}[htbp]
\begin{center}
\includegraphics[width=75mm]{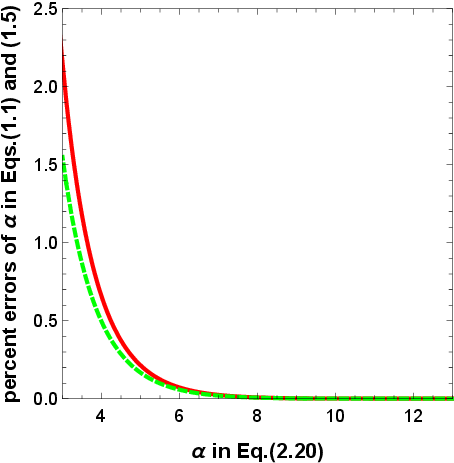}
\end{center}
\caption{The percent errors of deflection angles in the Ellis-Bronnikov wormhole spacetime. 
Solid (red) and dashed (green) curves denote
the percent errors of the deflection angles in Eqs.~(\ref{eq:alp11}) and (\ref{eq:alp12}), respectively, 
against the deflection angle in Eq.~(\ref{eq:alpha}).}
\label{fig:errorEllis}
\end{figure}
The percent errors of the deflection angles~(\ref{eq:alp11}) and~(\ref{eq:alp12}) are shown in Fig.~\ref{fig:errorEllis}.

\section{Gravitational lensing in usual lens configuration}
\begin{figure}[htbp]
\begin{center}
\includegraphics[width=60mm]{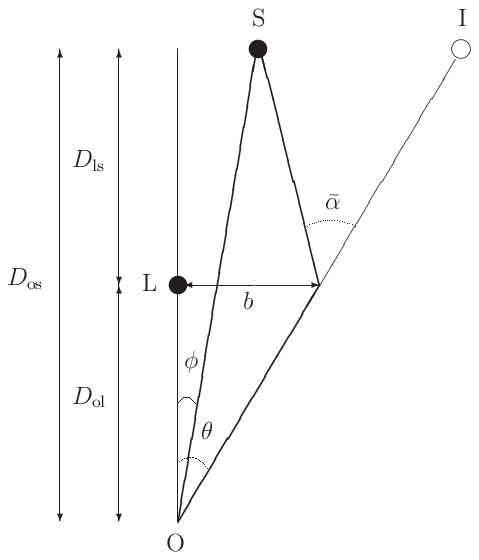}
\end{center}
\caption{Configuration of gravitational lensing. 
A ray with an impact parameter  $b$ is emitted by a source S with a source angle $\phi$, it is reflected with an effective deflection angle $\bar{\alpha}$
by a lens object L, and it is observed by an observer O as an image~I with an image angle $\theta$.
$D_{\mathrm{os}}$, $D_{\mathrm{ls}}$, and $D_{\mathrm{ol}}=D_{\mathrm{os}}-D_{\mathrm{ls}}$ denote distances between O and S, between L and S, and between O and L, respectively. 
}
\label{fig:conf}
\end{figure}

We consider that a ray with an impact parameter $b$, which is emitted by a source S with a source angle $\phi$, 
is deflected with a deflection angle $\alpha$ by a lens object L and its image I with an image angle $\theta$ is observed by an observer O as shown in Fig.~\ref{fig:conf}.
The distances between O and S, between L and S, and between O and L are denoted by $D_{\mathrm{os}}$, $D_{\mathrm{ls}}$, and $D_{\mathrm{ol}}=D_{\mathrm{os}}-D_{\mathrm{ls}}$, respectively.
By using an effective deflection angle $\bar{\alpha}$ defined by
\begin{eqnarray}
\bar{\alpha}= \alpha \quad  \mathrm{mod} \quad 2\pi,
\end{eqnarray}
a small-angle lens equation \cite{Bozza:2008ev} is expressed by 
\begin{eqnarray}\label{eq:lens}
D_{\mathrm{ls}} \bar{\alpha}= D_{\mathrm{os}}(\theta-\phi),
\end{eqnarray}
where we have assumed $\left| \bar{\alpha} \right| \ll 1$, $\left| \theta \right| = \left| b  \right| /D_{\mathrm{ol}} \ll 1$, and $\left| \phi \right| \ll 1$.
The deflection angle $\alpha$ can be expressed by
\begin{eqnarray}\label{eq:def2}
\alpha=\bar{\alpha} + 2\pi N,
\end{eqnarray}
where $N$ is a winding number of the ray.
We define an angle $\theta^0_N$ by
\begin{eqnarray}\label{eq:def3}
\alpha(\theta^0_N)=2\pi N
\end{eqnarray}
and we expand the deflection angle $\alpha(\theta)$ around $\theta=\theta^0_N$ as 
\begin{equation}\label{eq:def4}
\alpha(\theta)=\alpha(\theta^0_N)+ \left. \frac{d \alpha}{d \theta} \right|_{\theta=\theta^0_N} (\theta-\theta^0_N) +O\left( \left( \theta-\theta^0_N \right)^2 \right).
\end{equation}

\subsection{By using the deflection angle~(\ref{eq:alp12})}
We express the $0$th order of affine perturbation series of the deflection angle~(\ref{eq:alp12}) as 
\begin{eqnarray}\label{eq:def1}
\alpha(\theta)
&=&-\bar{a} \log \left( 1-\frac{\theta_\infty}{\theta} \right) +\bar{b} \nonumber\\
&&+O\left( \left( 1-\frac{\theta_\infty}{\theta} \right) \log \left( 1-\frac{\theta_\infty}{\theta} \right) \right),
\end{eqnarray}
where $\theta_\infty \equiv b_\mathrm{m}/D_{\mathrm{ol}}$ is the image angle of the photon sphere.
From Eqs.~(\ref{eq:def3}) and (\ref{eq:def1}), we get 
\begin{equation}\label{eq:theta0N}
\theta^0_N=\frac{\theta_\infty}{1-e^\frac{\bar{b}-2\pi N}{\bar{a}}}.
\end{equation}
From 
\begin{equation}
\left. \frac{d \alpha}{d \theta} \right|_{\theta=\theta^0_N}=\frac{\bar{a}\theta_\infty}{\theta^0_N(\theta_\infty-\theta^0_N)},
\end{equation}
and Eqs.~(\ref{eq:def2})-(\ref{eq:def4}) and (\ref{eq:theta0N}), the effective deflection angle $\bar{\alpha}(\theta_N)$, where $\theta=\theta_N$ is the positive solution of the lens equation for a positive winding number $N$,
is obtained as 
\begin{equation}\label{eq:effdef}
\bar{\alpha}(\theta_N)=-\frac{\bar{a}\theta_\infty (\theta_N-\theta_N^0)}{\left(\theta_N^0 \right)^2 e^\frac{\bar{b}-2\pi N}{\bar{a}}}.
\end{equation}
By substituting the effective deflection angle (\ref{eq:effdef}) to the lens equation (\ref{eq:lens}), we obtain the image angle as 
\begin{equation}
\theta_N (\phi) \sim \theta_N^0 +\frac{\left(\theta_N^0 \right)^2 e^\frac{\bar{b}-2\pi N}{\bar{a}}D_{\mathrm{os}}}{\bar{a}\theta_\infty D_{\mathrm{ls}}} \left( \phi-\theta_N^0 \right)
\end{equation}
and the image angle of an Einstein ring with the winding number $N$ as 
\begin{equation}
\theta_{EN} \equiv \theta_N (0) \sim \theta_N^0 \left( 1- \frac{\left(\theta_N^0 \right)^2 e^\frac{\bar{b}-2\pi N}{\bar{a}}D_{\mathrm{os}}}{\bar{a}\theta_\infty D_{\mathrm{ls}}}  \right).
\end{equation}
The difference of image angles between the outermost image and the photon sphere is given by
\begin{equation}
\bar{s}\equiv \theta_1 -\theta_\infty \sim \theta_1^0-\theta_\infty^0 =\frac{\theta_\infty e^\frac{\bar{b}-2\pi}{\bar{a}}}{1-e^\frac{\bar{b}-2\pi}{\bar{a}}}.
\end{equation}

The magnification of the image is given by 
\begin{equation}\label{eq:mag}
\mu_N(\phi)
\equiv \frac{\theta_N}{\phi} \frac{d \theta_N}{d \phi}
\sim \frac{ \theta_\infty^2 e^\frac{\bar{b}-2\pi N}{\bar{a}}D_{\mathrm{os}}}{\phi \bar{a} D_{\mathrm{ls}} \left( 1-e^\frac{\bar{b}-2\pi N}{\bar{a}} \right)^3 }.
\end{equation}
The ratio of the magnifications of the outermost image to the sum of the other images is obtained by 
\begin{equation}
\bar{r}\equiv \frac{\mu_1}{\sum^\infty_{N=2} \mu_N} \sim \frac{e^\frac{\bar{b}-2\pi}{\bar{a}}}{\left( 1-e^\frac{\bar{b}-2\pi}{\bar{a}} \right)^3\sum^\infty_{N=2} \frac{e^\frac{\bar{b}-2\pi N}{\bar{a}}}{\left( 1-e^\frac{\bar{b}-2\pi N}{\bar{a}} \right)^3}} .
\end{equation}

Notice that we can get a negative solution $\theta=\theta_{-N}(\phi)\sim -\theta_N(\phi)$ of the lens equation for each winding number $N$ 
while we have concentrated on the positive solution $\theta=\theta_N(\phi)$. 
The separation of the positive and negative image angles for each $N$ is given by $\theta_N(\phi)-\theta_{-N}(\phi)\sim 2\theta_N(\phi)$.
The magnification of the negative image angle $\theta_{-N}(\phi)$ is given by $\mu_{-N}(\phi)\sim -\mu_{N}(\phi)$.
The total magnification $\mu_{N\mathrm{tot}}(\phi)$ of the positive and negative image angles for each $N$ is given by
\begin{equation}
\mu_{N\mathrm{tot}}(\phi)
\equiv \left| \mu_{N}(\phi) \right| + \left| \mu_{-N}(\phi) \right| 
\sim \frac{2 \theta_\infty^2 e^\frac{\bar{b}-2\pi N}{\bar{a}}D_{\mathrm{os}}}{\phi \bar{a} D_{\mathrm{ls}} \left( 1-e^\frac{\bar{b}-2\pi N}{\bar{a}} \right)^3 }.
\end{equation}
Table~I shows the observables with the deflection angle~(\ref{eq:alp12}). 
\begin{table*}[htbp]
 \caption{Gravitational lensing in a usual lens configuration with the deflection angle~(\ref{eq:alp12}): 
 $\bar{a}$, $\bar{b}$, $2\theta_{\infty}$, $2\theta_{\mathrm{E}1}$, $\bar{\mathrm{s}}$, $\mu_{1\mathrm{tot}}(\phi)$, 
 and $\bar{\mathrm{r}}$ in the Schwarzschild and Reissner-Nordstr\"om black hole spacetimes for given $Q/M$ 
 and in the Ellis-Bronnikov wormhole spacetime are shown.
 We set the mass $M=6.5\times 10^9 M_{\odot}$, distances $D_{\mathrm{os}}=33.6$~Mpc
 and $D_{\mathrm{ol}}=D_{\mathrm{ls}}=16.8$~Mpc, 
 and the source angle $\phi=1$ arcsecond. 
 The parameter $a=4 \sqrt{2/\pi} \left( D_{\mathrm{ls}}D_{\mathrm{ol}}/D_{\mathrm{os}} \right)^\frac{1}{4} M^\frac{3}{4}$ is set so that $2\theta_{\mathrm{E}0}$ of the Ellis-Bronnikov wormhole is the same value 
 as $2\theta_{\mathrm{E}0}=2.52$arcsecond of the Schwarzschild and Reissner-Nordstr\"om black holes.   
 }
\begin{center}
\begin{tabular}{ c c c c c c c c |c} \hline\hline
$Q/M$          	                   	 &&$0$        &$0.2$    &$0.4$     &$0.6$     &$0.8$    &$1$     &Wormhole \\ \hline
$\bar{a}$        	            	 &&$1.0000$     &$1.0046$   &$1.0197$    &$1.0518$    &$1.1232$   &$1.4142$   &$1.0000$   \\ 
$\bar{b}$        	            	 &&$-0.4002$   &$-0.3993$ &$-0.3972$  &$-0.3965$  &$-0.4136$ &$-0.7332$ &$-1.0622$ \\ 
$2\theta_{\infty}$~($\mu$as)	         &&$39.9132$    &$39.6450$  &$38.8135$   &$37.3211$   &$34.9191$  &$30.7252$  &$9923.91$  \\ 
$2\theta_{\mathrm{E}1}$~($\mu$as)        &&$39.9633$    &$39.6963$  &$38.8690$   &$37.3864$   &$35.0092$  &$30.9419$  &$9930.32$   \\ 
$\bar{\mathrm{s}}$~($\mu$as)             &&$0.02501$     &$0.02563$   &$0.02776$    &$0.03263$    &$0.04505$   &$0.1084$   &$3.205$    \\ 
$\mu_{1\mathrm{tot}}(\phi)\times10^{17}$ &&$0.9702$      &$0.9832$    &$1.0276$     &$1.1267$     &$1.3651$    &$2.3149$    &$30885$     \\ 
$\bar{\mathrm{r}}$                	 &&$536.5$       &$521.5$     &$475.2$     &$394.0$      &$269.9$     &$85.8$     &$535.5$     \\ 
\hline\hline
\end{tabular}
\end{center}
\end{table*}

\subsection{By using the deflection angle~(\ref{eq:alp11})}
As a reference, we consider the deflection angle~(\ref{eq:alp11}) in the strong deflection limit. It is rewritten in
\begin{eqnarray}\label{eq:def1-}
\alpha(\theta)
&=&-\bar{a} \log \left( \frac{\theta}{\theta_\infty}-1 \right) +\bar{b}.
\end{eqnarray}
By using Eqs.~(\ref{eq:def3}) and (\ref{eq:def1-}), we obtain 
\begin{equation}\label{eq:theta0N-}
\theta^0_N= \theta_\infty \left( 1+e^\frac{\bar{b}-2\pi N}{\bar{a}} \right).
\end{equation}
From 
\begin{equation}
\left. \frac{d \alpha}{d \theta} \right|_{\theta=\theta^0_N}=\frac{\bar{a}}{\theta_\infty-\theta^0_N},
\end{equation}
and Eqs.~(\ref{eq:def2})-(\ref{eq:def4}) and (\ref{eq:theta0N-}), we obtain the effective deflection angle $\bar{\alpha}(\theta_N)$ as
\begin{equation}\label{eq:effdef-}
\bar{\alpha}(\theta_N)=-\frac{\bar{a}(\theta_N-\theta_N^0)}{\theta_\infty e^\frac{\bar{b}-2\pi N}{\bar{a}}}.
\end{equation}
From Eqs. (\ref{eq:lens}) and  (\ref{eq:effdef-}), we get the image angle  
\begin{equation}
\theta_N (\phi) \sim \theta_N^0 +\frac{\theta_\infty e^\frac{\bar{b}-2\pi N}{\bar{a}}D_{\mathrm{os}}}{\bar{a} D_{\mathrm{ls}}} \left( \phi-\theta_N^0 \right),
\end{equation}
the image angle of an Einstein ring for each $N$ 
\begin{equation}
\theta_{EN} \sim \theta_N^0 \left(1- \frac{\theta_\infty e^\frac{\bar{b}-2\pi N}{\bar{a}}D_{\mathrm{os}}}{\bar{a} D_{\mathrm{ls}}} \right),
\end{equation}
and the difference of image angles between the outermost image and the photon sphere 
\begin{equation}
\bar{s} \sim \theta_1^0-\theta_\infty^0 =\theta_\infty e^\frac{\bar{b}-2\pi}{\bar{a}}.
\end{equation}

The magnification of the image is obtained as
\begin{equation}\label{eq:mag-}
\mu_N(\phi)
\sim \frac{ \theta_\infty^2 e^\frac{\bar{b}-2\pi N}{\bar{a}}D_{\mathrm{os}}  \left( 1+e^\frac{\bar{b}-2\pi N}{\bar{a}} \right) }{\phi \bar{a} D_{\mathrm{ls}} }
\end{equation}
and the ratio of the magnifications of the outermost image to the sum of the other images is given by
\begin{equation}
\bar{r} \sim \frac{\left(e^\frac{4\pi}{\bar{a}} -1\right)\left(e^\frac{2\pi}{\bar{a}} +e^\frac{\bar{b}}{\bar{a}}\right)}{e^\frac{4\pi}{\bar{a}}+e^\frac{2\pi}{\bar{a}}+e^\frac{\bar{b}}{\bar{a}}}.
\end{equation}

The separation of two images are obtained as $2\theta_N(\phi)$ and 
their total magnification for each $N$ is given by
\begin{equation}
\mu_{N\mathrm{tot}}(\phi)
\sim \frac{ 2 \theta_\infty^2 e^\frac{\bar{b}-2\pi N}{\bar{a}}D_{\mathrm{os}}  \left( 1+e^\frac{\bar{b}-2\pi N}{\bar{a}} \right) }{\phi \bar{a} D_{\mathrm{ls}} }.
\end{equation}
The observables with the deflection angle~(\ref{eq:alp11}) are shown in Table II.
\begin{table*}[htbp]
 \caption{Gravitational lensing in the usual lens configuration with deflection angle~(\ref{eq:alp11}): We set $M$, $D_{\mathrm{os}}$, $D_{\mathrm{ol}}$, $D_{\mathrm{ls}}$, 
$\phi$, and $a$ to be the same values as the ones in TABLE~I. We do not show $\bar{a}$, $\bar{b}$, and $2\theta_{\infty}$ 
because they give the same values as the ones in TABLE~I. 
 }
\begin{center}
\begin{tabular}{ c c c c c c c c |c} \hline\hline
$Q/M$          	                   	 &&$0$        &$0.2$    &$0.4$     &$0.6$     &$0.8$    &$1$     &Wormhole \\ \hline
$2\theta_{\mathrm{E}1}$~($\mu$as)        &&$39.9632$    &$39.6962$  &$38.8689$   &$37.3863$   &$35.0090$  &$30.9404$  &$9930.31$   \\ 
$\bar{\mathrm{s}}$~($\mu$as)             &&$0.02498$     &$0.02559$   &$0.02772$    &$0.03258$    &$0.04493$   &$0.1076$   &$3.203$    \\ 
$\mu_{1\mathrm{tot}}(\phi)\times10^{17}$ &&$0.9678$      &$0.9807$    &$1.0247$     &$1.1228$     &$1.3581$    &$2.2824$    &$30845$     \\ 
$\bar{\mathrm{r}}$                	 &&$535.2$       &$520.1$     &$473.9$     &$392.6$      &$268.5$     &$84.6$     &$534.8$     \\ 
\hline\hline
\end{tabular}
\end{center}
\end{table*}

\section{Retro lensing}
Gravitational lensing with the deflection angle $\alpha \sim \pi$ is called retro lensing.
Retro lensing in black hole spacetimes~\cite{Holz:2002uf,DePaolis:2003ad,Eiroa:2003jf,Bozza:2004kq,DePaolis:2004xe,Abdujabbarov:2017pfw,Tsukamoto:2016oca}, 
wormhole spacetimes~\cite{Tsukamoto:2017edq,Tsukamoto:2016zdu},
naked singularity spacetimes~\cite{ZamanBabar:2021zuk,Tsukamoto:2021lpm},
and black bounce spacetimes~\cite{Guerrero:2021ues,Tsukamoto:2022vkt}
were investigated.
In this section, we investigate retro lensing with a configuration that
a lens object L with a photon sphere, an observer O, and a source S are almost aligned in the order as shown in Fig.~4.%
\begin{figure}[htbp]
\begin{center}
\includegraphics[width=80mm]{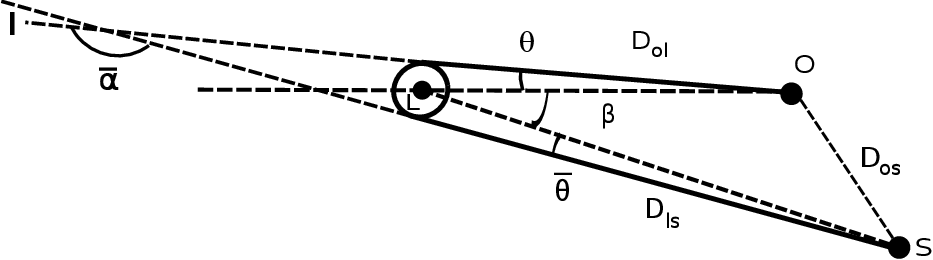}
\end{center}
\caption{Configuration of retro lensing. 
A ray emitted by a source S is reflected with an effective deflection angle $\bar{\alpha}$
by the photon sphere of a lens object L and it reaches to an observer O.
We introduce an source angle $\beta\equiv \angle$OLS and an angle $\bar{\theta}$ defined by an angle between a line LS and the ray at S.
}
\label{fig:retro}
\end{figure}
A light ray emitted by the source is reflected by the photon sphere of the lens object and it is observed by the observer as an image I.
The Ohanian lens equation~\cite{Ohanian_1987,Bozza:2008ev} is expressed as 
\begin{equation}\label{eq:retrolens}
\beta=\pi-\bar{\alpha}(\theta)+\theta+\bar{\theta},
\end{equation}
where $\beta \sim 0$ is a source angle defined by $\angle$OLS and $\bar{\theta}$ is an angle between a line LS and the light ray at S.
We also assume that the terms $\theta=b/D_{\mathrm{ol}}$ and $\bar{\theta}=b/D_{\mathrm{ls}}$ are small and that they can be neglected in the lens equation.
We obtain a positive solution $\theta=\theta_N$ for every winding number $N$ and its magnification is obtained as~\cite{Eiroa:2003jf,Bozza:2004kq,Tsukamoto:2016oca}
\begin{equation}
\mu_N (\beta)=- \frac{D^2_{\mathrm{os}}}{D^2_{\mathrm{ls}}} s(\beta) \theta_N \frac{d\theta_N}{d \beta},
\end{equation}
where $s(\beta)$ for a point source is given by
\begin{equation}
s(\beta)=\frac{1}{\beta}
\end{equation}
and for an uniform-luminous disk with a size $\beta_\mathrm{s} \equiv R_\mathrm{s}/D_{\mathrm{ls}}$, where $R_\mathrm{s}$ is the radius of the source, on a source plane, 
\begin{equation}
s(\beta)=\frac{1}{\pi \beta_\mathrm{s}^2} \int_\mathrm{Disk} d\beta^\prime d \Phi,
\end{equation}
where $\beta^\prime$ is a radial coordinate divided by $D_{\mathrm{ls}}$ on the source plane, $\Phi$ is an azimuthal coordinate 
around the origin of the coordinates on the source plane.
By fixing the point of an intersection between an axis $\beta=0$ and the source plane as the origin of the coordinates, 
$s(\beta)$ is given by 
\begin{equation}
s(\beta)=\frac{2}{\pi \beta_\mathrm{s}^2} \left[ \pi \left( \beta_\mathrm{s}-\beta \right) + \int^{\beta+\beta_\mathrm{s}}_{-\beta+\beta_\mathrm{s}} \arccos \frac{\beta^2+\beta^{\prime 2}-\beta_\mathrm{s}^2}{2 \beta \beta^\prime} d\beta^\prime \right] \nonumber\\
\end{equation}
for $\beta \leq \beta_\mathrm{s}$
and 
\begin{equation}
s(\beta)=\frac{2}{\pi \beta_\mathrm{s}^2} \int^{\beta+\beta_\mathrm{s}}_{\beta-\beta_\mathrm{s}} \arccos \frac{\beta^2+\beta^{\prime 2}-\beta_\mathrm{s}^2}{2 \beta \beta^\prime} d\beta^\prime
\end{equation}
for $\beta>\beta_\mathrm{s}$.
A perfectly-aligned case $\beta=0$ for the uniform-luminous disk with the finite size, we obtain 
\begin{equation}
s(0)=\frac{2}{\beta_\mathrm{s}}.
\end{equation}

\subsection{By using the deflection angle~(\ref{eq:alp12})}
We investigate retro lensing by using the deflection angle (\ref{eq:alp12}) or (\ref{eq:def1}).
From Eqs.~(\ref{eq:def2}), (\ref{eq:def1}), and (\ref{eq:retrolens}), 
we obtain the image angle $\theta_{N}$ with the winding number $N$ as 
\begin{equation}
\theta_N (\beta)=\frac{\theta_\infty}{1-e^\frac{\beta+\bar{b}-\pi(1+2N)}{\bar{a}}}
\end{equation}
and its magnification as 
\begin{equation}
\mu_N (\beta) =- \frac{D_{\mathrm{os}}^2}{D_{\mathrm{ls}}^2} \frac{\theta_\infty^2 e^\frac{\beta+\bar{b}-\pi(1+2N)}{\bar{a}} }{ \bar{a} \left(1-e^\frac{\beta+\bar{b}-\pi(1+2N)}{\bar{a}}\right)^3} s(\beta).
\end{equation}
We also get a negative image angle as $\theta_{-N} (\beta)\sim -\theta_N (\beta)$ and their total magnification as  
\begin{equation}
\mu_{N\mathrm{tot}}(\beta) \sim \frac{2D_{\mathrm{os}}^2}{D_{\mathrm{ls}}^2} \frac{\theta_\infty^2 e^\frac{\beta+\bar{b}-\pi(1+2N)}{\bar{a}} }{ \bar{a} \left(1-e^\frac{\beta+\bar{b}-\pi(1+2N)}{\bar{a}}\right)^3} s(\beta).
\end{equation}
Table~III shows observables by retro lensing with the deflection angle~(\ref{eq:alp12}). 
\begin{table*}[htbp]
 \caption{Retro lensing with deflection angle~(\ref{eq:alp12}): 
 The total magnification $\mu_{0\mathrm{tot}}(0)$ and apparent magnitude of the sun lensed by a photon sphere at $D_{\mathrm{ol}}=0.01$pc 
 We have assumed $M=60 M_{\odot}$ for the Schwarzschild and Reissner-Nordstr\"om black hole
 and $a=4 \sqrt{2/\pi} \left( D_{\mathrm{ls}}D_{\mathrm{ol}}/D_{\mathrm{os}} \right)^\frac{1}{4} M^\frac{3}{4}$ for the Ellis-Bronnikov wormhole.
 }
\begin{center}
\begin{tabular}{ c c c c c c c c |c} \hline\hline
$Q/M$          	                       &&$0$        &$0.2$    &$0.4$     &$0.6$     &$0.8$    &$1$        &Wormhole \\ \hline
$\mu_{0\mathrm{tot}}(0)\times10^{20}$  &&$2.967$    &$2.968$  &$2.975$   &$2.999$   &$3.070$  &$3.101$    &$1.478\times 10^6$     \\ \hline  
apparent magnitude                     &&$22.030$   &$22.030$ &$22.027$  &$22.019$  &$21.993$ &$21.982$   &$7.7865$     \\
\hline\hline
\end{tabular}
\end{center}
\end{table*}

\subsection{By using the deflection angle~(\ref{eq:alp11})}
As a reference, we consider retro lensing with the deflection angle~(\ref{eq:alp11}). 
From Eqs.~(\ref{eq:def2}), (\ref{eq:def1-}), and (\ref{eq:retrolens}), 
we obtain the image angle $\theta_{N}$ with the winding number $N$ as 
\begin{equation}
\theta_N (\beta)=\theta_\infty \left(1+e^\frac{\beta+\bar{b}-\pi(1+2N)}{\bar{a}}\right)
\end{equation}
and its magnification as 
\begin{equation}
\mu_N (\beta) =- \frac{D_{\mathrm{os}}^2}{D_{\mathrm{ls}}^2} \frac{\theta_\infty^2 e^\frac{\beta+\bar{b}-\pi(1+2N)}{\bar{a}}  \left(1+e^\frac{\beta+\bar{b}-\pi(1+2N)}{\bar{a}}\right) }{ \bar{a}} s(\beta).
\end{equation}
The total magnification of the positive and negative images for the winding number $N$ is obtained as  
\begin{equation}
\mu_{N\mathrm{tot}}(\beta) \sim \frac{2D_{\mathrm{os}}^2}{D_{\mathrm{ls}}^2} \frac{\theta_\infty^2 e^\frac{\beta+\bar{b}-\pi(1+2N)}{\bar{a}} }{ \bar{a} \left(1-e^\frac{\beta+\bar{b}-\pi(1+2N)}{\bar{a}}\right)^3} s(\beta).
\end{equation}
The observables by retro lensing with the deflection angle~(\ref{eq:alp11}) are shown in Table~IV. 
\begin{table*}[htbp]
 \caption{Retro lensing with deflection angle~(\ref{eq:alp11}):  We have assumed the same parameters as Table~III.
 }
\begin{center}
\begin{tabular}{ c c c c c c c c |c} \hline\hline
$Q/M$          	                       &&$0$        &$0.2$    &$0.4$     &$0.6$     &$0.8$    &$1$        &Wormhole \\ \hline
$\mu_{0\mathrm{tot}}(0)\times10^{20}$  &&$2.795$    &$2.793$  &$2.790$   &$2.791$   &$2.811$  &$2.702$    &$1.434\times 10^6$     \\ \hline
apparent magnitude                     &&$22.095$   &$22.096$ &$22.097$  &$22.097$  &$22.089$ &$22.132$   &$7.8194$     \\ 
\hline\hline
\end{tabular}
\end{center}
\end{table*}

\section{Conclusion and discussion}
We have shown the $0$th order of affine perturbation series of the deflection angle~(\ref{eq:alp12}) 
is more accurate than the deflection angle~(\ref{eq:alp11}), which is often used in the strong deflection limit $b\rightarrow b_\mathrm{m}+0$, 
not only in the black hole spacetimes but also in the wormhole spacetime. 
We have investigated gravitational lensing by using the $0$th order of affine perturbation series of the deflection angle~(\ref{eq:alp12}). 
As shown Tables~I and II, under the usual lens configuration with the photon sphere, 
the observables obtained by using the $0$th order of the affine perturbation series of the deflection angle~(\ref{eq:alp12}) and  
by using the deflection angle~(\ref{eq:alp11}) in the strong deflection limit are almost the same.
Thus, we can ignore the effect of the hidden errors in the deflection angle~(\ref{eq:alp11}) in the strong deflection limit
on the observables in the usual lens configuration.

On the other hand, in a retro lensing configuration, the error of the deflection angle~(\ref{eq:alp11}) in the strong-deflection-limit analysis increases to be several percents 
and the error of the $0$th order of affine perturbation series of the deflection angle~(\ref{eq:alp12}) reduces to be the almost half of the error of the deflection angle~(\ref{eq:alp11}).
We have shown that the hidden errors in the deflection angle~(\ref{eq:alp11}) can affect on the magnification by a dozen percent. 
Thus, we conclude that we should use the $0$th order of affine perturbation series of the deflection angle ~(\ref{eq:alp12}) rather than the deflection angle~(\ref{eq:alp11}) when we consider retro lensing.
The $0$th order of affine perturbation series of the deflection angle can give a brighter magnification by a dozen percent than 
the one by using the deflection angle~(\ref{eq:alp11}) in the strong-deflection-limit analysis.

On this paper, we have investigated gravitational lensing by a photon sphere in  a general, asymptotically flat, static, and spherically symmetric spacetime 
with a photon sphere. One might think that some numerical treatments of gravitational lensing in given spacetimes
are more practical than analytical treatments and one does not need analytical studies.
However, we can reveal universal property of the gravitational lensing in the general spacetime 
with the photon sphere by using the analytical method while we have to rely on concrete examples by numerical methods.
Therefore, it is still meaningful to consider the analytical treatment even if some examples were calculated numerically.

Gravitational lensing by rotating black holes in strong deflection limit has been investigated 
in Refs.~\cite{Bozza:2005tg,Bozza:2002af,Bozza:2004kq,Bozza:2005tg,Bozza:2006nm,Hsieh:2021scb,Ghosh:2022mka}.
On this paper, we have concentrated on asymptotically flat, static, and spherically symmetric spacetimes
and our results can be extended to the rotating black holes and alternatives. 
The extension to the rotating case is left as future work.

%
\appendix
\section{Weak-field approximations}
In this appendix, we review gravitational lensing under weak-field approximations in the usual lens configuration.

\subsection{Schwarzschild and Reissner-Nordstr\"om spacetimes} 
Under the weak-field approximation $\left| b \right| \gg M$ in the Schwarzschild and Reissner-Nordstr\"om spacetimes,
the deflection angle (\ref{eq:alpha}) can be expressed by 
\begin{equation}
\alpha=\frac{4M}{b}.
\end{equation}
By using the deflection angle and Eqs.~(\ref{eq:lens}) and (\ref{eq:def2}),
$\theta=b/D_{\mathrm{ol}}$, and $N=0$, we get the reduced lens equation 
\begin{equation}
\hat{\theta}^2-\hat{\phi}\hat{\theta}-1=0,
\end{equation}
and its solutions as 
\begin{equation}
\hat{\theta} = \hat{\theta}_{\pm 0}(\hat{\phi}) \equiv \frac{\hat{\phi}\pm \sqrt{\hat{\phi}^2+4}}{2},
\end{equation}
where $\hat{\theta}\equiv \theta/\theta_{\mathrm{E}0}$ and $\hat{\phi}\equiv \phi/\theta_{\mathrm{E}0}$ are a reduced image angle and a reduced source angle,
respectively, and $\theta_{\mathrm{E}0}$ is the image angle of an Einstein ring given by 
\begin{equation}
\theta_{\mathrm{E}0}= 2\sqrt{\frac{MD_{\mathrm{ls}}}{D_{\mathrm{ol}}D_{\mathrm{os}}}}.
\end{equation}
The magnifications of the images and its total magnification are obtained by
\begin{eqnarray}
\mu_{\pm 0}
&\equiv&\frac{\hat{\theta}_{\pm 0}}{\hat{\phi}} \frac{d\hat{\theta}_{\pm 0}}{d\hat{\phi}} \nonumber\\
&=& \frac{1}{4} \left( 2 \pm \frac{\hat{\phi}}{\sqrt{\hat{\phi}^2+4}} \pm \frac{\sqrt{\hat{\phi}^2+4}}{\hat{\phi}} \right)\nonumber\\
&=& \frac{\hat{\theta}_{\pm 0}^4}{\left( \hat{\theta}_{\pm 0}^2 \mp 1 \right) \left( \hat{\theta}_{\pm 0}^2 \pm 1 \right)}
\end{eqnarray}
and 
\begin{eqnarray}
\mu_{0\mathrm{tot}} 
&\equiv& \left| \mu_{+0} \right| +\left| \mu_{-0} \right| \nonumber\\
&=& \frac{1}{2} \left( \frac{\hat{\phi}}{\sqrt{\hat{\phi}^2+4}} + \frac{\sqrt{\hat{\phi}^2+4}}{\hat{\phi}} \right),
\end{eqnarray}
respectively.

\subsection{Ellis-Bronnikov wormhole spacetime} 
Under the weak-field approximation $\left| b \right| \gg a$ in the Ellis-Bronnikov wormhole spacetime~\cite{Abe:2010ap,Toki:2011zu,Kitamura:2012zy,Tsukamoto:2012zz,Takahashi:2013jqa,Yoo:2013cia,Izumi:2013tya,Nakajima:2014nba,Bozza:2015haa,Bozza:2015wbw,Bozza:2017dkv,Tsukamoto:2017hva,Bozza:2020ubm},  
the deflection angle (\ref{eq:alpha}) is given by 
\begin{equation}\label{eq:Ealpha}
\alpha=\pm \frac{\pi a^2}{4b^2}.
\end{equation}
From  Eqs.~(\ref{eq:lens}), (\ref{eq:def2}), and (\ref{eq:Ealpha}),
$\theta=b/D_{\mathrm{ol}}$, and $N=0$, we get the reduced lens equation 
\begin{equation}
\hat{\theta}^3-\hat{\phi}\hat{\theta}^2\mp1=0
\end{equation}
and the image angle of the Einstein ring is obtained as
\begin{equation}
\theta_{\mathrm{E}0}= \left( \frac{\pi a^2D_{\mathrm{ls}}}{4 D_{\mathrm{ol}}^2D_{\mathrm{os}}}\right)^\frac{1}{3}.
\end{equation}
The lens equation always has a positive solution $\hat{\theta}=\hat{\theta}_{+0}(\hat{\phi})$ and a negative one $\hat{\theta}=\hat{\theta}_{-0}(\hat{\phi})$ 
and their magnifications are expressed by
\begin{eqnarray}
\mu_{\pm 0}= \frac{\hat{\theta}_{\pm 0}^6}{\left( \hat{\theta}_{\pm 0}^3 \mp 1 \right) \left( \hat{\theta}_{\pm 0}^3 \pm 2 \right)}.
\end{eqnarray}

In Tables~I-IV, we set parameter $a$ as
\begin{eqnarray}
a=4 \sqrt{\frac{2}{\pi}} \left( \frac{D_{\mathrm{ls}}D_{\mathrm{ol}}}{D_{\mathrm{os}}} \right)^\frac{1}{4} M^\frac{3}{4}
\end{eqnarray}
so that $\theta_{\mathrm{E}0}$ in the Ellis-Bronnikov wormhole spacetime is the same 
as $\theta_{\mathrm{E}0}$ in the Schwarzschild and Reissner-Nordstr\"om spacetimes.

\end{document}